\documentstyle[12pt,leng]{article}
\begin{document}
{\flushright IP-ASTP-32-93}\\
{\flushright January 12, 1994}
\vspace{24pt}
\begin{center}
{\large\sc{\bf Landau Level Gap Reduction at Abrupt Edges.}}
\baselineskip=12pt
\vspace{35pt}

I. Barto\v{s}\footnote{permanent address: Institute of Physics,
Academy of Sciences of the Czech Republic, Prague} and B.Rosenstein
\vspace{24pt}

 Institute of Physics \\
Academia Sinica\\
Taipei, 11529\\
Taiwan\\
\vspace{260pt}
\end{center}
\baselineskip=24pt
PACS numbers: 73.20 Dx, 73.40 Hm, 03.65 Ge.
\vspace{30pt}
\pagebreak

\begin{center}
{\bf abstract}
\end{center}

For an electron localized near a finite  rectangular step potential
under strong magnetic field we found a profound local reduction of the gap
between neighbouring Landau levels \cite{BR,BJ}. We investigate
under what conditions the effect persists when the barrier gets
smoother. The conclusion is that  to get a substantial
gap reduction the barrier gradient has to be
larger than $\hbar\omega_c/a_L$ ($a_L$ is the magnetic length).
The sensitivity of the gap reduction is illustrated using
tilted and multiple step barriers and finite stripe configurations.
\vspace{10pt}

\pagebreak
{\bf Introduction.}\newline

In the  modern edge state picture of the magnetotransport phenomena in
two dimensional electron gas (2DEG)
in a strong perpendicular magnetic field \cite{H,Ha,K} all the current
flows through one dimensional channels located in the vicinity of
the system edges. Therefore  magnetotransport is very sensitive to
the electron structure in regions close
to effective confinement barriers of 2DEG.
In an attempt to treat more realistically the shape of the barrier
we replaced the widely used infinite barrier \cite{MDS} by a finite
rectangular barrier \cite{BR} (to be referred as I).

In the model with a finite rectangular potential barrier confining
2DEG in a strong magnetic field we observed a steplike
spectrum of electron energies. Landau level energies,
which depend on the distance from the barrier,
exhibit flat plateaus. There is a curious "apparent level crossing":
a higher level approaches very closely the lower one.  Neighbouring
Landau levels have been found to approach each other locally reducing
bulk energy gap between them $\hbar\omega_c$ to just a few
percents. Simultaneously, the space  separation between
neighbouring edge channels gets
reduced. This results in strong enhancement  of the coupling
among the channels which leads to corresponding changes in magnetotransport.

This contrasts with the standard assumption that energies of all the
edge states rise smoothly.
It is indeed the case in two limiting situations widely
discussed in the literature: a very smooth
onset of the potential barrier \cite{Ha} and in the infinite
abrupt step model \cite{MDS}.
In the first case, Landau levels simply bend following the shape of
the underlying smooth potential and remain parallel. In the second,
they rise steeply approaching the infinite barrier but still keep the
energy separation between the levels intact.
The phenomenon of gap reduction will profoundly
influence the transport properties and density of  edge states
of the sample. This, in turn, can be observed, for example, in nuclear
magnetic resonance in these systems.

In I we considered  an abrupt relatively
large potential barriers  $V$ few times
higher than the inter  - Landau level spacing, $\hbar\omega_c$. The main
purpose of the present paper is to study how general the gap
reduction is. In order to learn more about the sensitivity of
 the phenomenon on the
steepness of the barrier we investigate more general types of
barriers.

As we
show in section 1, the same phenomenon, though less pronounced,
 occurs also for
the physically important barrier of the height
$V=\hbar \omega_c$.
 In section 2 we smooth the barrier by tilting the potential wall,
while in the section 3 a single abrupt step is replaced by a sequence
of smaller rectangular steps. Recent self consistent calculations
for confinement barriers with various degrees of
steepness \cite{C,BPT} have shown that the effective potential
develops steps  with the unit height $\hbar\omega_c$ .
Therefore we investigate in detail the electron
structure in the vicinity of the unit height barriers.
Although both types of barriers in magnetic field
can still be treated analytically as in I, more general approach will
be adopted here.
In section 4 similar gap reductions are observed for a finite stripe
configurations. The limiting case of narriw stripe representing a line
impurity is discussed in section 5. We conclude with brief discussion
of possible experiments revealing the the gap reduction.\newline

{\bf 1. The unit rectangular barrier.}\newline

As in I, we investigate a semiinfinite 2DEG, confined in the $x$
direction by a barrier situated at $x=0$.
In Fig. 1, the electron structure in the vicinity of the rectangular step
barrier of height $V=1$  in units of $\hbar\omega_c$ is shown.
In contrast to cases with slowly varying (on the scale of the
magnetic length) confining  potentials in which the electron energies simply
follow the potential shape, $E_n(X)$ behave differently. Here $X$
is the distance of the center of the Larmor orbit from the edge.
The energy separation between the two lowest Landau levels does
change. As the Larmor orbit center $X$ approaches the
barrier (situated at $x=0$), energy of the higher
Landau level starts rising earlier
 due to its larger extent of the wave function than for
the lower state. This rise saturates for the higher level, exhibiting
a plateau at $X>0$ because of the node
in its center. At the end of this saturation $X\sim 1$ (in units of
 magnetic length), the separation
between the two branches is significantly
reduced. As the asymptotic region
is approached,  the bulk gap value gets gradually recovered.
 Though the gap reduction is not as
large as
that for higher barrier in I it is still significant, representing
 reduction of about 20$\%$.\newline

{\bf 2. Tilted barrier.}\newline

Sensitivity of the gap narrowing to the steepness of the
confining potential barrier will be tested here in a model with
a linearly rising onset of the potential barrier V(x).
\begin{equation}
V(x)=\cases{0 & for $x<0 $\cr $(V/W) \ \ x $ &
 for $0<x<W$ \cr V & for $x>W$\cr}
\end{equation}
with a single parameter $W$ for the range at which the potential
reaches its maximal value $V=1$ in units of $\hbar\omega_c$.

Following standard steps (see, for example, \cite{MDS}),
in the Landau gauge  the effective
one dimensional equation for an electron in magnetic field becomes
\begin{equation}
\left[-\frac{\hbar}{2m}\frac{d^2}{dx^2}+\frac{1}{2}m\omega_c^2(x-X)^2+
V(x)\right]
\phi_{n,X}(x)=E_{n,X}\phi_{n,X}(x)
\end{equation}
The eigenvalue problem has been solved by a numerical integration
from one asymptotic region  ($X\rightarrow -\infty$) over the
confinement barrier range to the
opposite asymptotic region ($X\rightarrow \infty$). The energy dependance
of the logarithmic derivatives of the solutions is very high ensuring
precise eigenvalue determination. This approach ,
in addition, can be used for any shape
of the confinement barrier.

Fig. 2a for $W=1$ (in units of magnetic length
$a_L\equiv{\sqrt {\hbar c/eB}}$) shows that the gap narrowing,
observed for a rectangular barrier
(Fig. 1 ) is basically preserved here. The reduction is still around 20\%.
On the other hand, on Fig. 2b and 2c for $W=2 $ and 4 the effect gradually
disappears. In the latter case the energy levels copy the underlying potential
shape. From this we conclude that the sharpness of the barrier of the
order of the Larmor orbit radius is needed to generate the quantum
mechanical phenomenon described above. For smoother potentials the
energy levels $E_n(X)$ follow the potential, as is
 also the case of the self consistent
calculation \cite{BPT}.\newline

{\bf 3. Double step barrier.}\newline

As in self consistent calculations \cite{C,BPT}, the effective potential
develops steps of Landau spacing in the vicinity of the system edge,
we study the double step  with various widths to investigate the
conditions under which the gap reduction persists.

Our model potential with a step of width $W$ is
\begin{equation}
V(x)=\cases{0 & for $x<0$\cr V & for $0<x<W$\cr 2V & for $x>W$\cr}
\end{equation}
If the width of the step is much larger than the magnetic length,
energy levels $E_n(X)$ are composed of two consecutive step barrier
profiles in the vicinity of each step as expected. See Fig. 3b for $W=4$.
When the step width is reduced to the order of magnitude of the magnetic
length, differences between the individual orbitals, resulting
from their different sizes, start to appear. This is illustrated
for $W=1$ in Fig. 3a: higher orbitals effectively average the
potential barrier and their energies rise gradually to their asymptotic values
in contrast to the lowest Landau orbital which exhibits characteristic
nonuniform features discussed above. As a result, substantial reduction
(about 30\%) of the gap between the two lowest branches is observed at
$X=2$.\newline

{\bf 4. The stripe configuration.}\newline

Similar behaviour, as in the semiinfinite 2DEG, is exhibited also in the
finite stripe configuration (see also \cite{BJ}).
 For symmetry reasons the center of the well
will be placed at $x=0$.
 Again, the gap reduction above the barrier tops
is found, see Fig. 4. In addition, as the stripe width  $W$
is gradually decreased,
the bound states are pushed to higher energies, as has been observed
in the model study of superlattices in
strong magnetic fields \cite{M}. The characteristic picture of oscillations
appears on Fig. 4. The intermediate  width $W=2$ and small width
$W=.5$ illustrate the trend (Fig. 4b and 4a respectively).
The limiting case of a very narrow stripe is the $\delta$ potential
which can be solved analytically \cite{JC}. \newline

{\bf 5. The $\delta$ potential.}\newline

The eigenvalues of electron in magnetic field and the line
potential $V(x,y)=a\delta (x)$ are obtained by solving the
transcendental equation \cite{JC}
\begin{equation}
\frac{D_{E+1}(X)}{D_E(X)}+\frac{D_{E+1}(-X)}{D_E(-X)}=-\sqrt{2} a
\end{equation}
where $D_E(X)$ are  the parabolic cylinder functions
\cite{E}. The results for both attractive ($a=-3$)
and repulsive potential ($a=3$) are shown in Fig. 5a and 5b respectively.
The only qualitative difference between the two is the existence
of a lone bound state for the attractive potential.
The parabolic shape of the deformed lowest Landau level is  understood
perturbatively \cite{A}.
The gap reductions are again observed for $a$ larger than product of
magnetic length and the Landau level spacing.

In analogy to finite barrier case, "almost level crossings" here
correspond to intersections of the parabolas with horizontal unperturbed
straight Landau levels. Oscillations with increasing numbers of
nodes appear in the continuum inside the parabola.

The line $\delta$ function perturbation of 2DEG in magnetic field
may represent following two physical
situations. The first is "lateral $\delta$ doping" within the 2DEG
plane with the magnetic field perpendicular to the plane.
Another is the usual $\delta$ doping in superlattices with
the magnetic field aligned
parallel to the superlattice layers.\newline

{\bf Discussion.}

To summarize, we qualified the types of barriers
for which the phenomenon of  local gap reduction between Landau levels  occurs.
The barrier should be higher than Landau level spacing and the gradient
should be larger than
\begin{equation}
\frac{\partial V}{\partial X}=\frac{\hbar\omega_c}{a_L}=\frac{\hbar^{1/2}}
{m}\left(\frac{eB}{c}\right)^{3/2}
\end{equation}
For magnetic field $B=20 \ \ T$ the Landau level spacing  in GaAs is
about $30 \ \ meV$ and
the magnetic length is $60  \ \ {\dot A}$ we get  $\frac{\partial V}{\partial
X}=
.5 \ \ \frac{meV}{{\dot A}}$.

Such gradients can be achieved in lateral
heterojunctions if, for example, the sample consists of a few segments
 of different semiconductors (e.g.  GaAs/AlGaAs).
These gradients
have to be present also in laterally $\delta$ doped samples.
 The localized defect,  here, by no means
 has to be as narrow as an interatomic distance to produce the
effect.

The presence of the gradients leads to a number of observable consequences.
First,
the reduction of energy gap obviously changes the coupling between
neighbouring edge channels.  As a consequence one expects qualitative changes
in resistivity as function of magnetic field . The change is rather
profound since the dependence of the coupling on the interchanell separation
is exponential (\cite{H} and references therein).

Second, the density of electron states as a function of
energy is qualitatively modified by the gap reduction effect. It should
 be  directly observable by means of the nuclear spin relaxation effect.
The nuclear spin
 relaxation  is due to the  contact coupling between the nuclear and
the electron spins. In 2DEG under strong magnetic field it
 was predicted \cite{V}
and subsequently observed \cite{BK}. The inverse relaxation time is proportial
to the density of the relevant electron states. Very recently it has been shown
 that the contribution
of the edge states for some magnetic field regions may be dominant \cite{VMS}.
The change in the density of states due to the platous inside the edge region
is expected to be observable in optical experiments.

In the calculations presented here, electron - electron interactions
are not taken into account. The selfconsistent treatment would modify
the effective potential. However for steep confinement barriers the
barrier effect itself dominates (see also \cite{DGH}). Our approach
is general enough to enable the self consistent treatment
of the system.

We acknowledge the support of  National Science Council ROC, grants
NSC-82-0208-M-001-116 (B.R.) and NSC-82-0501-I-00101-B11 (I.B.).

\newpage

FIGURE CAPTIONS\newline
\newline

Fig. 1

{\it Unit rectangular step barrier.}\newline
Eigenvalues of two lowest Landau levels of the semiinfinite 2DEG
confined by the  rectangular step barrier of height one indicated by the dashed
line.  The energies are functions of $X$, the Larmor's orbit
center with respect to the step barrier
. Energies are given in units of
Landau level spacing $\hbar\omega_c$ and distances in units of magnetic length
${\sqrt \frac{\hbar c}{eB}}$.
\newline\newline

Fig. 2

{\it Tilted barriers.}\newline
Eigenvalues of four lowest
 Landau levels in the vicinity of the
potential barrier.  The confining barrier rises
linearly over the region of width $W$ to its final value of 1, see eq.(1).
a) $W=.5$, b) $W=1$, c) $W=2$. Units are  the same as in Fig. 1.
\newline\newline

Fig. 3

{\it Double step barriers.}\newline
Eigenvalues of four lowest
 Landau levels in the vicinity of the
potential barrier.  The confining barrier rises
in two steps of equal height
 to its final value of 1. The width of the first step is $W$ ,
see eq.(3).
a) $W=1$, b) $W=4$. Units are the same as in Fig. 1.
\newpage

Fig. 4

{\it Stripe configurations.}\newline
Eigenvalues of lowest
 Landau levels for a stripe, infinite in $y$ direction and finite (width $W$)
in the
$x$ direction. Rectangular step barriers of unit height confine the electron
gas. Due to symmetry only the right half of the stripe is shown.
The eigenstates are pushed up to higher energies as the stripe width
decreases: a) $W=1$, b) $W=2$. Units are the same as in Fig. 1.

Fig. 5

{\it $\delta$ function potential.}\newline
Eigenvalues of lowest
 Landau levels for an infinitely narrow stripe (situated at $x=0$
and infinite in $y$ direction), determined by eq.(4).
The strength of the $\delta$ potential is $a=3$.
Units are same as in Fig. 1. $X$ is the distance of the center of the Larmor
orbit from the $\delta$ function. Results for both an attractive
and a repulsive cases are shown in Fig. (a) and (b) respectively. Due to the
symmetry only the left part of the picture is shown.

\end{document}